\documentclass[12pt]{iopart}
\usepackage{iopams}
\usepackage{graphicx}
\usepackage{color}

\newcommand{\veca}{\mathbf{a}}

\newcommand{\vecE}{\mathbf{E}}
\newcommand{\vecB}{\mathbf{B}}

\newcommand{\vecF}{\mathbf{F}}

\newcommand{\vecv}{\mathbf{v}}

\newcommand{\beq}{\begin{equation}}
\newcommand{\eeq}{\end{equation}}

\begin{document}
\title
[Efficiency of radiation friction losses in laser-driven ``hole boring'']
{
Efficiency of radiation friction losses in laser-driven ``hole boring'' of dense targets
}

\author{S V Popruzhenko$^{1,2}$, T V Liseykina$^{3,4}$, A Macchi$^{5,6}$}
\address{$^1$Max Planck Institute for the Physics of Complex Systems, Dresden 01187, Germany\\
$^2$National Research Nuclear University MEPhI, Moscow 115409, Russia\\
$^3$Institute of Physics, University of Rostock, D-18051 Rostock, Germany\\
$^4$On leave from Institute of Computational Mathematics and Mathematical Geophysics SB RAS, 630090 Novosibirsk, Russia\\
$^5$ CNR/INO (National Institute of Optics), Adriano Gozzini unit, 56124 Pisa, Italy\\
$^6$ Enrico Fermi Department of Physics, University of Pisa, largo Bruno Pontecorvo 3, 56127 Pisa, Italy}
\ead{sergey.popruzhenko@gmail.com}

\begin{abstract}
In the interaction of laser pulses of extreme intensity ($>10^{23}~{\rm W cm}^{-2}$) with high-density, thick plasma targets, simulations show significant radiation friction losses, in contrast to thin targets for which such losses are negligible. 
We present an analytical calculation, 
based on classical radiation friction modeling, of the conversion efficiency of the laser energy into incoherent radiation 
in the case when a circularly polarized pulse interacts with a thick plasma slab of overcritical initial density. 
By accounting for three effects including the influence of radiation losses on the single electron trajectory, the global ``hole boring'' motion of the laser-plasma interaction region under the action of radiation pressure, and the inhomogeneity of the laser field in both longitudinal and transverse direction, we find a good agreement with the results of three-dimensional particle-in-cell simulations.
Overall, the collective effects greatly reduce radiation losses with respect to electrons driven by the same laser pulse in vacuum, which also shift the reliability of classical calculations up to higher intensities.
\end{abstract}

\section{Introduction}\label{sec:intro}

The continuous progress of laser techniques making higher and higher electromagnetic intensities accessible for experiments has stimulated the growth of research areas such as relativistic dynamics and nonlinear optics in classical plasmas \cite{bulanov-rmp06} and quantum electrodynamics in extremely strong fields \cite{dipiazza-rmp12,narozhny-cp15}. Radiation friction (RF) is a problem of central interest in both the above mentioned areas. In the classical context, a modification of the Newton-Lorentz equation of motion for an electron by adding a new force term, named the RF force (RFF) or radiation reaction force, is necessary to make the electron dynamics self-consistent with the emission of radiation. Although the correct form of the RFF has been the subject of intense debate for decades and until recently \cite{sokolov-jetp09,zotev-pop16}, it now appears that in the classical limit the Landau-Lifshitz (LL) expression \cite{landau-lifshitz-RR} gives a correct and consistent description \cite{krivitskii-spu91,spohn-el00}. The LL expression of the RFF has become the basis of classical simulations of superintense laser-plasma interaction \cite{tamburini-njp10,vranic-cpc16} where RF losses (corresponding to the escape of high-frequency, incoherent radiation from the plasma) are important enough to affect the plasma dynamics. 

When the frequency of the emitted radiation becomes sufficiently high that the energy and momentum of single photons are not negligible with respect to those of the radiating electron, a quantum electrodynamics (QED) description becomes necessary. 
However, a correct and effective description of ``quantum RF'' is an open issue. 
The first two experiments claiming for evidence of quantum RF signatures in nonlinear Thomson scattering of superintense laser pulses by ultrarelativistic electrons \cite{cole-prx18,poder-prx17} came to somewhat different conclusions about which model better described the experimental results (see \cite{macchi-p18} for a discussion).  
Notice that these experiments involved laser-plasma physics in the generation via wakefield acceleration of a dense, short duration bunch of relativistic electrons in order to increase the luminosity in the gamma-ray region; however, the dynamics of the laser-bunch interaction was of single particle nature.
The geometry of these experiments was designed to maximize RF losses in order to make quantum signatures appearing at relatively low intensity. 
In this regime, such signatures are mostly a reduction of RF losses with respect to the classical calculation because of the spectral cut-off which appears when the emitted photon energy approaches the photon energy. 
These effects can be reproduced by a semiclassical modeling, similar to what found in a different class of experiments involving high energy electron scattering in crystals \cite{andersen-prd12,wistisenNC18}.

An alternative approach to investigate RF in the laboratory is to search for regimes where \emph{collective} effects in the laser-plasma interaction boost radiation losses, so that RF signatures may become strong and unambiguous. 
Several simulation works have shown highly efficient radiation losses (a few tens per cent of the laser pulse energy) in the interaction of circularly polarized (CP) pulses with dense thick targets \cite{naumova-prl09,schlegel-pop09,capdessus-pop14,capdessus-pre15,nerush-ppcf15,ournjp,delsorbo-njp18}. 
This is in sharp contrast with thin targets accelerated by CP pulses in the so-called ``light sail'' (LS) regime, for which the radiation losses are very weak \cite{tamburini-njp10,tamburiniPRE12,zhangNJP15}. 
Strong differences between CP and linear polarization (LP) were also evidenced \cite{tamburini-njp10,tamburiniPRE12}. Hence, the collective laser-plasma dynamics can play a crucial role in determining the amount of RF losses.

In our previous work \cite{ournjp} we have made a first attempt of a classical model to estimate the conversion efficiency $\eta_{\rm rad}$ of the laser energy into incoherent radiation in the case when a strong CP pulse interacts with a thick plasma of overcritical initial density.
In turn, the efficient absorption of CP light causes a strong transfer of angular momentum to the target, with the generation of ultrahigh magnetic fields (inverse Faraday effect) with strength achieving several Giga-Gauss which can provide a macroscopic signature of RF \cite{ournjp}. 

In Ref.\cite{ournjp} the scaling of $\eta_{\rm rad}$ with the laser intensity agreed reasonably with the results of three-dimensional particle-in-cell (PIC) simulations of the laser-plasma interaction, up to intensities approaching $10^{24}~{\rm W cm}^{-2}$. Beyond this limit, however, the model predicts unphysical values of $\eta_{\rm rad}>1$ because neither the modification of the radiating electron trajectories due to RF nor the depletion of the laser pulse are taken into account. In addition, and more in general, at those intensities the classical description becomes questionable and quantum effects are expected to become relevant. 

The aim of this paper is to provide an accurate estimate of $\eta_{\rm rad}$ for CP fields via analytical modeling assuming that the classical RF regime is retained.
First, we use the solution by Zeldovich \cite{zeld} to take self-consistently into account the RF losses into the electron trajectory. 
Then, we show that the amount of RF losses is strongly affected by the average motion of the plasma surface, the finite evanescence length of the electromagnetic (EM) field in the plasma, and the radially inhomogeneous distribution of the laser intensity. 
By accounting for these effects, analytical estimates in good agreement with the results of three-dimensional (3D) simulations are obtained. 
We also provide an estimate for the value of the quantum parameter and show that, in the present context, the electron dynamics can still be well described within the classical RFF approach.

\section{Review of previous modeling and its limitations}\label{sec:review}

In the regime of interest here, an ultraintense laser pulse of frequency $\omega$ and dimensionless field amplitude $a_0=eE_L/m_e\omega c$ (with $E_L$ the electric field amplitude) interacts with a strongly overdense (electron density $n_e\gg n_c=m_e\omega^2/4\pi e^2$, the cut-off density) plasma target which remains opaque to the laser light. 
The radiation pressure of the laser light is high enough to produce ``hole boring '' (HB) in the target, i.e. the plasma surface is driven at an average velocity
\beq
\frac{v_{\rm HB}}{c}=\frac{\sqrt{\Xi}}{1+\sqrt{\Xi}} \; , \qquad \Xi=\frac{I_L}{\rho c^3}=\frac{Zn_cm_e}{An_em_p}a_0^2 \; ,
\label{V-hb}
\eeq
where $I_L=cE_L^2/4\pi=m_ec^3n_ca_0^2$ is the laser intensity.
Eq.(\ref{V-hb}) can be obtained by balancing the mass and momentum flows at the surface \cite{robinson} and is valid for total reflection of the laser light \emph{in the frame co-moving with the surface}, i.e. in the absence of dissipative effects. If a fraction $\eta$ of the laser intensity is dissipated, for example due to RF losses, Eq.(\ref{V-hb}) may be modified by replacing $I_L$ with $I_L(1-\eta/2)$. In the case of our simulations this would lead at most by a $\simeq 5\%$ decrease in $v_{\rm HB}$ at the highest intensity considered ($a_0=800$).

In order for the interaction to remain in the HB regime during the whole duration of the laser pulse, the target must be ``thick'' enough that $v_{\rm HB}\tau_L<D$, where $\tau_L$ is the laser pulse duration and $D$ is the target thickness. 
In the opposite ``thin'' target limit $v_{\rm HB}\tau_L\gg D$, the target can be accelerated as a whole and enter the ``light sail'' regime \cite{esirkepovPRL04,macchiNJP10}, where the scaling of the velocity $v_{\rm LS}$ with intensity becomes much faster than (\ref{V-hb}).
Thus, the same laser pulse parameters may enable to reach velocities $v_{\rm LS} \simeq c$ in an ultrathin target while yielding $v_{\rm HB}$ to be a fraction of $c$ in a thick target. 
In particular, for the parameters of calculations presented below $v_{\rm HB}\approx(0.3\div 0.6)c$.
The different acceleration regime may therefore explain the huge difference in the radiation efficiency between thick and thin targets. In fact, assuming that the electrons radiate in the field of a plane electromagnetic CP wave propagating along $x$, 
the radiated power is \cite{landau-lifshitz-Prad}
\beq
P_{\rm rad}=\frac{2e^2\omega^2\gamma^2a_0^2}{3c}\bigg(1-\frac{v_x}{c}\bigg)^2 \; .
\label{P-1}
\eeq  
where $\gamma=1/\sqrt{1-v^2/c^2}$ and $v_x$ is the velocity component parallel to the wavevector.
Assuming that most of the radiating electrons co-move with the target ions, the factor $(1-{v_x}/{c})^2$ leads to strong suppression of radiation emission for thin targets moving at $v_{\rm LS} \simeq c$, while the suppression is much less severe for thick targets as far as $v_{\rm HB}$ is significantly smaller than $c$. 

In the thick target case, the laser field penetrates into the skin layer where the electrons pile up under the action of the radiation pressure. The areal density of electrons in the skin layer can be estimated as \cite{ournjp,macchi-prl05} 
\beq
N_x\simeq \frac{a_0}{r_0\lambda}~,
\label{Nx}
\eeq
where $r_0=e^2/mc^2$ is the classical electron radius.
For $a_0\gg 1$, by estimating $\gamma \simeq a_0$ we obtain the radiated power per unit surface as $I_{\rm rad}=N_xP_{\rm rad} \propto a_0^5$, which implies a $\propto a_0^3$ scaling for the radiation loss efficiency, in good agreement with the simulation results. 
For $v_x=0$, and assuming that the duration of the uncoherent high-energy emission is the same as the laser pulse,    
the conversion efficiency defined as a ratio of the energy emitted by radiating electrons $U_{\rm rad}$ to that of the laser pulse $U_{\rm L}$ is thus given by 
\beq
\eta_{\rm rad}=\frac{U_{\rm rad}}{U_{\rm L}}=\frac{I_{\rm rad}}{I_{\rm L}}=\xi a_0^3~,
\label{eta-xi}
\eeq
where the parameter 
\beq
\xi=\frac{4\pi r_0}{3\lambda}
\label{xi}
\eeq
is introduced, and $\lambda$ is the laser wavelength. 

In Ref.\cite{ournjp} it was suggested that for thick targets an enhancement of radiation losses may originate from the non-steady dynamics of HB acceleration \cite{macchi-prl05}. In particular,
ion acceleration by the space-charge field causes a pulsed ``collapse'' of the electron density with the excess electrons returning towards the laser with \emph{negative} velocity $v_x<0$, enhancing the RF losses by a sequence of radiation bursts. 
However, it is not straightforward to provide analytical estimates for either the rate of the bursts or the value of $v_x$ for the returning electrons. 
In particular, estimating $v_x$ would require to find the motion of the returning electrons in an inhomogeneous electric field with the radiation friction force included. 
For an order-of-magnitude estimate, we simply assumed $(1-v_x/c)^2\simeq 1$ and the number of the returning electrons to be $\simeq N_x$ \cite{ournjp} (i.e. most of the electrons in the skin layer to collapse). This leads again to an expression like (\ref{eta-xi}) for the conversion efficiency, apart from a reduction factor $<1$ accounting for the fact that the returning electrons radiate only for a fraction of the interaction time.

Apparently, the $\eta_{\rm rad} \sim a_0^3$ scaling fairly agrees with the results of 3D simulations which give $\eta_{\rm rad}\sim a_0^{3.2}$ up to intensities $a_0\simeq 500$, but the absolute value predictions of (\ref{eta-xi}) are much higher than those observed in the simulations 
(see Fig.1 below). 
This is not surprising since obviously (\ref{eta-xi}) becomes invalid when approaching a critical value of the laser field amplitude
\beq
a_{\rm cr}=\xi^{-1/3}\approx 400~\; ,
\label{acr}
\eeq
where $\eta_{\rm rad}\simeq 1$, which is unphysical. 
In the simulation \cite{ournjp}, $\eta_{\rm rad}\approx 0.08$ for $a_0\simeq 400$. 
The quantitative disagreement makes also not possible, on the basis of the predicted scaling only, to understand whether the radiation is mostly due to electrons either remaining in the skin layer or returning towards the laser.

The very limited nature of the estimate (\ref{eta-xi}) for the conversion efficiency is due to several underlying shortcomings, such as the neglect of self-consistent RF effects on the electron motion, the absence of a more precise estimate of $v_x$, and the inhomogeneity of the laser field in both the longitudinal and transverse directions.
In the following we show that accounting for these effects, even if still in an approximated way, leads to a considerably smaller growth of the conversion efficiency at high intensities than that given by (\ref{eta-xi}) and therefore substantially improves the agreement with the simulations.  

\section{Self-consistent electron motion}

The model first introduced by Zeldovich \cite{zeld} describes a stationary electron motion in the field of a strong circularly polarized plane wave with RF effects included self-consistently. Since RF allows absorption of momentum from the plane wave, a drag force is exerted on the electron along the direction of wave propagation ($x$ for definiteness). Thus, in order to obtain a stationary solution an electric field $E_d$ along $x$ is introduced in the model balancing the radiation drag.
The complete EM fields are thus given by
\beq
\vecE(t,x)=(E_d,E_L\cos\varphi, E_L\sin\varphi)~,~~~~\vecB(t,x)=(0,-E_L\sin\varphi, E_L\cos\varphi)~.
\label{EH}
\eeq
In the stationary regime an electron moves along a circle in the $(y,z)$ plane and drifts along the $x$ axis with a constant velocity:
\beq
\vecv(t)=(v_x,v_0\sin(\varphi-\theta),-v_0\cos(\varphi-\theta))~,~~~~~\varphi=\omega t-kx~.
\label{v}
\eeq
The phase shift $\theta$ is generated by the RFF. 
Neglecting the latter in the equations of motion gives $\theta=0$.
For ultra-relativistic particles, the RFF is given by \cite{landau-lifshitz-RR}
\beq
\vecF_{\rm rad}=-\frac{P_{\rm rad}(\vecv)\vecv}{c^2}~,
\label{Fr}
\eeq
with the radiation power
\beq
P_{\rm rad}(\vecv)=\frac{2e^2\omega^2v_0^2\gamma^4}{3c^3}\left(1-\frac{v_x}{c}\right)^2
\label{P-20}
\eeq
which differs from (\ref{P-1}) by the replacement $a_0^2\to\gamma^2$ reflecting the fact that the circular motion of the electron is now determined jointly by the Lorentz and the RF forces.
In the stationary regime the total force (with the centrifugal component included) vanishes.
Projecting this condition on the axes of cylindrical coordinates and  assuming that the value of the longitudinal electric field $E_d$ is known  we obtain three equations which determine the values of $\gamma$, $\theta$ and $v_x:$ 
\beq
eE_d-\frac{eE_Lv_0}{c}\sin\theta-P_{\rm rad}(\gamma,v_x)\frac{v_x}{c^2}=0~,
\label{z}
\eeq
\beq
eE_L\bigg(1-\frac{v_x}{c}\bigg)\sin\theta+P_{\rm rad}(\gamma,v_x)\frac{v_0}{c^2}=0~,
\label{long}
\eeq
\beq
\gamma m\omega v_0=eE_L\cos\theta
\label{perp}
\eeq
In principle the system of Eqs.(\ref{z}-\ref{long}-\ref{perp}) might be applied to study the motion of  electrons in the space-charge field created by the ponderomotive force action, see examples e.g. in \cite{kost-pop16} where an approximate analytic description for the case of standing waves was developed.
However, such space-charge field is highly inhomogeneous, which would already make an analytical estimate difficult. In addition, in the case under investigation the electron density is high enough for screening effects to be non-negligible: considering as an example the contribution of returning electrons, as those located exactly at the plasma--vacuum boundary return towards the incoming laser, the space-charge field is partially canceled so that the electrons filling in inner layers will experience a lower force. A complete description of this scenario would require to resolve the electron plasma dynamics with RFF included. 

Since our primary aim is to relate the radiation losses to an average value of $v_x$ determined by the laser-plasma dynamics, we take $v_x$ as a parameter in the system, and following Zeldovich \cite{zeld} we solve Eqs.(\ref{z})--(\ref{perp}) in the reference frame moving with the instant velocity $v_x$ of the radiating electron. 
In the following, we use the notations $\gamma^{\prime}$, $v_0^{\prime}$, $\xi^{\prime}$, etc. for values measured in this reference frame.
Setting $v_x^{\prime}=0$ and taking into account that $\gamma^{\prime}\gg 1$, one may safely put $v_0^{\prime}\approx c$.
Eliminating the angle $\theta^{\prime}$ from Eqs.(\ref{long}), (\ref{perp}) we obtain an equation determining $\gamma^{\prime}(\xi^{\prime},a_0)$ (note that $a_0$ is relativistically invariant) \cite{zeld}:
\beq
{\gamma^{\prime}}^2( 1+{\xi^{\prime}}^2{\gamma^{\prime}}^6)=a_0^2~.
\label{ga0}
\eeq
For low intensities, $a_0\ll a_{\rm cr}^{\prime}$ it gives $\gamma^{\prime}=a_0$, as was used in \cite{ournjp}.
In the opposite limit, $a_0\gg a_{\rm cr}^{\prime}$, the gamma-factor grows much slower with $a_0:$
\beq
\gamma^{\prime}=\bigg(\frac{a_0}{\xi^{\prime}}\bigg)^{1/4}~.
\label{gamma-new}
\eeq
Eq.(\ref{gamma-new}), previously obtained in Ref.\cite{bulanovjrPRL10}, corresponds to the limit in which the oscillation energy of the electron in the EM field equals the energy radiated per cycle. 
Remarkably, this single particle result corresponds, in our model where collective effects enter via (\ref{Nx}) for the number of radiating electrons, to a total conversion of the laser pulse energy into radiation from the target. 
In fact it follows from Eqs.(\ref{ga0}), (\ref{P-20}) and (\ref{Nx}) that
\beq
\eta_{\rm rad}=\xi^{\prime}\frac{{\gamma^{\prime}}^4}{a_0}
=\xi\sqrt{\frac{1-v_x/c}{1+v_x/c}}\frac{{\gamma^{\prime}}^4}{a_0}~,
\label{eta-new}
\eeq
so that $\eta_{\rm rad}\to 1$ for $a_0\gg a_{\rm cr}^{\prime}$. 
Note that $\xi$ is determined by the laser wavelength measured in the laboratory frame.

\begin{figure}
\centerline{
\includegraphics[width=0.8\textwidth]{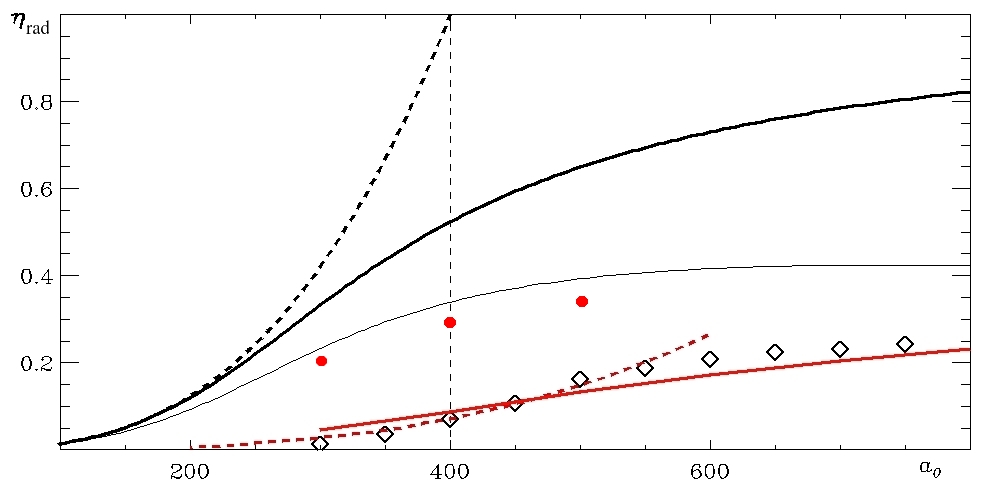}
}
\caption{
Conversion efficiency into radiation $\eta_{\rm rad}$ as a function of the dimensionless laser amplitude $a_0$. 
The dashed line shows the perturbative result (\ref{eta-xi}) \cite{ournjp}.
Results of the selfconsistent calculation are shown by a thick black line (Eq.(\ref{eta-new}) with $v_x=0$), thin black line (Eq.(\ref{eta-new}) with $v_x=v_{\rm HB}$ from (\ref{V-hb})) and thick red line (Eq.(\ref{eta-5}) with $v_x=v_{\rm HB}$).
Results of 3D PIC simulations are shown by empty black diamonds and filled red circles for circular and linear polarization of the laser pulse, respectively. 
The dashed red line is a $\sim a_0^{3.2}$ fit to the PIC results for circular polarization.
}  
\label{fig1} 
\end{figure}

We compare predictions of our model to the results of 3D PIC simulations (see \cite{ournjp} for the numerical set-up details) which describe the interaction of a laser pulse with a plasma of thickness $D>10\lambda$ 
where $\lambda=0.8\mu$m corresponding to a Ti:Sapphire laser and initial density $n_0=90n_c=1.55\cdot 10^{23}{\rm cm}^{-3}.$ 
The charge-to mass ratio for ions was taken $Z/A=1/2$.
The supergaussian laser pulse is introduced via the time-dependent boundary condition at the plasma surface, $x=0$, as described in \cite{ournjp}
\beq
\veca(r,x=0,t)=a_0({\bf y}\cos(\omega t)+{\bf z}\sin(\omega t))\mbox{e}^{-(r/r_0)^4-(ct/r_L)^4}~,
\label{a-rxt}
\eeq
with $r=\sqrt{y^2+z^2}$, $r_0=3.8\lambda$, $r_L=3.0\lambda$ and duration 
(full-width-half-maximum of the intensity profile) 14.6 fs.
In our PIC calculations we varied the laser amplitude in the interval $a_0=300\div 750$ which corresponds to the peak intensities $(3.8\div 23.7)\cdot 10^{23}$W/cm$^2$ 
and the total pulse energy (1.08--6.71)kJ.
The numerical box had a ~$[30\times 25\times 25]\lambda^3$ size, with 40 grid cells per $\lambda$ in each direction and 125 particles per cell 
for each species. The simulations were performed on $5000\div 10000$ cores of the JURECA Cluster Module at NIC (J\"ulich, Germany).

As is seen on Fig.\ref{fig1}, the values of $\eta_{\rm rad}$ obtained from (\ref{eta-new}) and shown by thick black line for $v_x=0$ qualitatively reproduce the behavior of conversion efficiency extracted from the PIC simulation (shown by diamonds) in the whole interval of $a_0$, although the absolute values appear considerably overestimated.
Below we identify the sources of these differences and improve the model by accounting for the respective effects.
For the sake of comparison, three values of $\eta_{\rm rad}$ are also shown for a LP pulse at all other parameters identical to those of the CP simulation.
As it is seen, the conversion efficiency is considerably higher in the LP case and also reaches its saturation at lower intensities. 
This is in line with previous observations \cite{tamburini-njp10,tamburiniPRE12} and can be traced back to effect of the magnetic (${\bf v}\times{\bf B}$) force driving longitudinal electron oscillations during which $v_x \rightarrow -c$.

\section{Effects of the longitudinal velocity}

\begin{figure}
\centerline{
\includegraphics[width=0.8\textwidth]{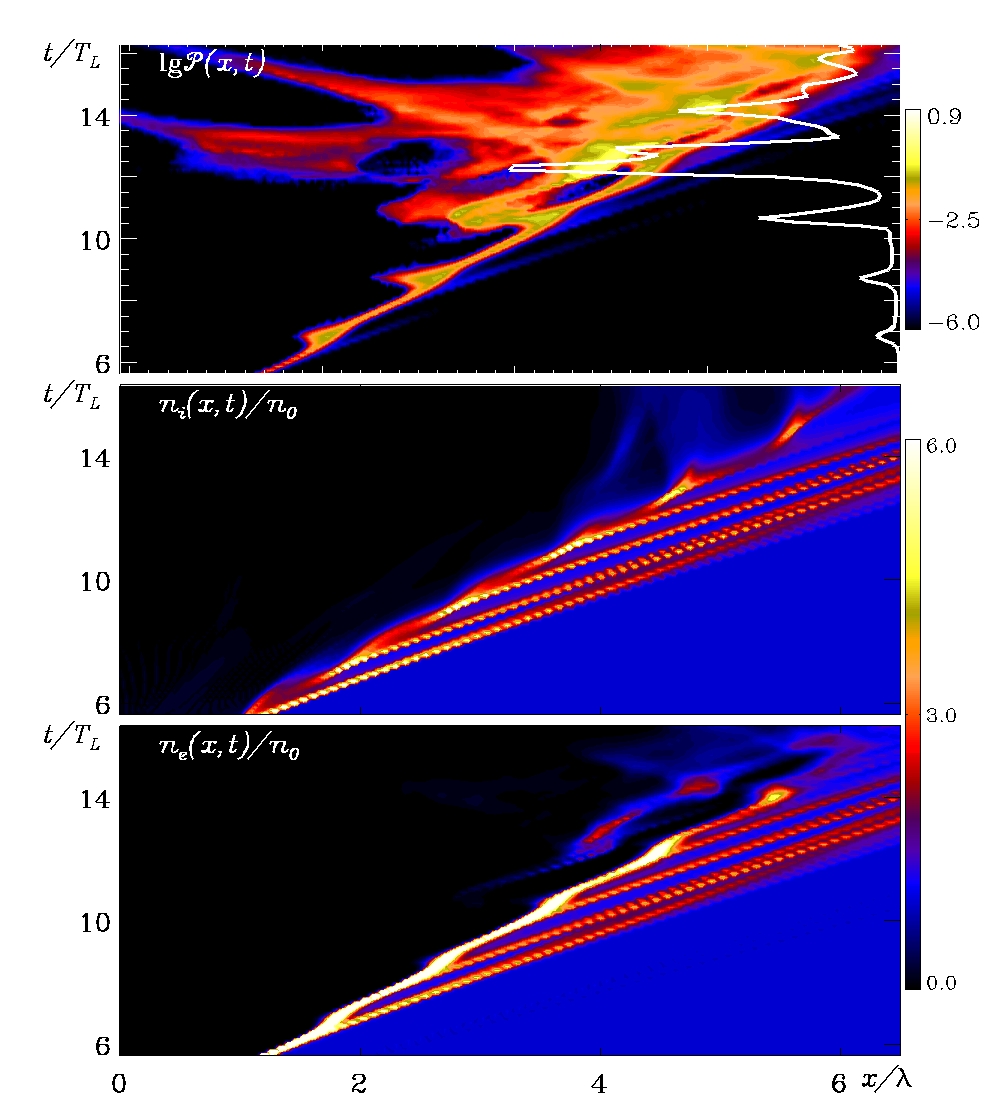}
}
\caption{Space-time plots of the radiation power density ${\cal P}(x,t)$ (top, logarithmic scale, arbitrary units), ion density $n_i(x,t)$ (middle) and electron density $n_e(x,t)$ (bottom) all evaluated at $r=1\lambda$ distance from the axis. 
A white curve on panel (a) shows the radiation power integrated over $x$.
}
\label{fig2} 
\end{figure}

An analysis of the 3D distribution functions of the radiation power density ${\cal P}(x,r,v_x)$ (calculated as ${\cal P}=-n_e\vecv\cdot\vecF_{\rm rad}$) and of the electron and ion density $n_{e,i}(x,r,v_x)$ extracted from the PIC simulation shows that most of the emitted radiation comes from electrons having velocities $v_x>0$, and located close to the receding front of the ion density.
This is illustrated for the $a_0=500$ case in Fig.\ref{fig2} where space-time plots in the $(x,t)$ plane are shown for the radiation power and the particle densities at $r=1\lambda$, where the former has its radial maximum. The density fronts move in the forward direction with average velocity $\simeq 0.41c$, in fair agreement with the value $v_{\rm HB}=0.47c$ given by Eq.(\ref{V-hb}). Small oscillations in the front position are visible in correspondence of the generation of plasma bunches in the forward direction, as discussed in Ref.\cite{macchi-prl05}. 
The power density plot shows that most of the emission originates close to the hole boring front. 
Emission due to returning electrons with velocity $\simeq -c$ is visible after $t=11T_L$, but its contribution to the total emitted power is small, presumably because of the low density in the returning jets (as seen the $n_e(x,t)$ plot).

As clearly seen from the plot of the $x$-integrated radiation power shown on panel (a), spikes of radiation occur in correspondance of the generation of plasma bunches. Such spikes may be explained by the enhanced penetration of the laser field into the plasma at these time instants. Since the spikes remain close to the hole boring front, no strong modification of $v_x$ is correlated with them.
Consistently with these observations, we assume that on the average the radiating electrons move with velocity $v_x=v_{\rm HB}$ given by (\ref{V-hb}).
In this way, we obtain a result shown on Fig.\ref{fig1} by a thin black line. 
The account of the longitudinal motion considerably improves the agreement, although analytically calculated values still exceed the PIC results by approximately 4 times at $a_0=400$ and 1.5 times at $a_0=750$.

\section{Effects of field inhomogeneity}\label{sec:inhom}

\begin{figure}
\centerline{\includegraphics[width=0.6\textwidth]{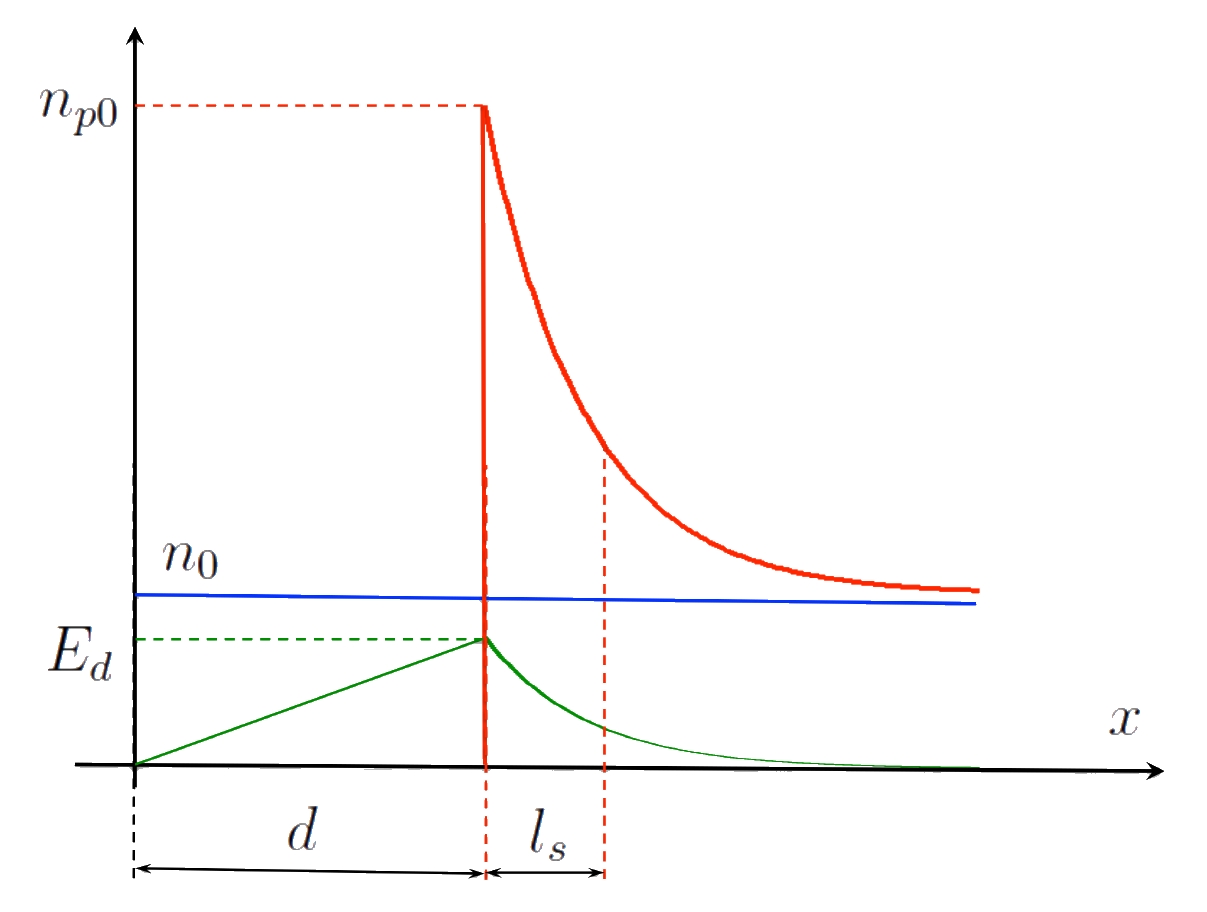}}
\caption{Distribution of electron (red line) and ion (blue line) charge densities calculated using the HB model for the initial stage of the charge separation, before the ions started moving under the action of the longitudinal electric field (green curve).}  
\label{fig3} 
\end{figure}

Finally, we account for the attenuation of the laser field in the plasma and the dependence of the laser intensity on time and its radial distribution in the focal spot.
The laser field amplitude $a_0$ is not constant within the evanescence length $\ell_s$, but dropping down, leading to a considerable decrease of the ``efficient'' value of $a_0$ entering Eqs.(\ref{gamma-new}) and (\ref{eta-new}).
Figure \ref{fig3} based on the HB model of \cite{macchi-prl05} sketches the electron and the ion density distributions along the propagation direction at the initial stage of the interaction when the electrons are pushed forward by light pressure, while the ions still remain immobile and homogeneously distributed inside the plasma layer.
Taking the electron density for $x>d$ in the form
\beq
n_e(x)=n_0+(n_{p0}-n_0)\mbox{e}^{-(x-d)/\ell_s}~,
\label{ne}
\eeq
we replace a step distribution employed in Ref.\cite{ournjp} by a decaying exponent.
Assuming that $n_{p0}\gg n_0$ with $n_{p0}$ being the maximal density of electrons and $n_0$ is the initial density equal to that of ions, we obtain for the electric field inside the layer
\beq
E(x)=4\pi e(n_{p0}-n_0)l_s\mbox{e}^{-(x-d)/\ell_s}
\label{Ex}
\eeq
with the maximal value 
\beq
E_d\equiv E(x=d)=4\pi e(n_{p0}-n_0)\ell_s\approx 4\pi en_{p0}\ell_s
\label{Ed}
\eeq
achieved at the electron surface.
Taking into account that $n_{p0}\ell_s= N_x\simeq a_0/r_0\lambda$ (\ref{Nx}) we obtain for the maximal longitudinal field
\beq
E_d\approx 3E_{\rm cl}\xi a_0~,
\label{Ed-1} 
\eeq
where $E_{\rm cl}=e/r_0^2=m^2c^4/e^3=1.81\cdot 10^{18}$V/cm is the critical field of classical electrodynamics which is $1/\alpha=137$ times greater than that of quantum electrodynamics 
\beq
E_{\rm cr}=\frac{m^2c^3}{e\hbar}~.
\label{Ecr}
\eeq
Note that for $\lambda\simeq 1\mu{\rm m}$, $E_d\simeq E_{\rm cr}$ at $a_0\simeq (400\xi)^{-1}\approx 1.6\cdot 10^5$, according to (\ref{Ed-1}), so that in this case $E_d\simeq E_L\simeq E_{\rm cr}$.

Within the same approximation the local equilibrium condition for the electrons inside the layer requires that the laser field amplitude drops accordingly, $a(x)=a_0\exp(-(x-d)/\ell_s)$.
Then the global equilibrium condition for the whole layer reads
\beq
(2-\eta_{\rm rad})\frac{I_L}{c}=e\int\limits_d^{\infty}n_e(x)E(x)dx\approx 2\pi e^2n_{p0}^2\ell_s^2~.
\label{equil}
\eeq
Here we take into account that the intensity of reflected radiation is $(1-\eta_{\rm rad})I_L$.
The arial radiation power (intensity) is
\beq
I_{\rm rad}^{\prime}=\int\limits_d^{\infty}P^{\prime}(x,v_x=0)n_e(x)dx=\frac{2e^2{\omega^{\prime}}^2a_0^4n_{p0}\ell_s}{3c}f(\xi^{\prime},a_0)~,
\label{W}
\eeq
where the power $P^{\prime}(x,v_x=0)$ is given by (\ref{P-20}) in the reference frame co-moving with the electrons and $\gamma^{\prime}(\xi^{\prime},a_0)$ is expressed from Eq.(\ref{ga0}), so that
\beq
f(\xi,a)=\frac{1}{a^5}\int\limits_0^a{\gamma}^4(\xi,\bar{a})d\bar{a}~.
\label{fa}
\eeq
This gives the following equation for the conversion efficiency 
\beq
\eta_{\rm rad}=\frac{I_{\rm rad}^{\prime}}{I_L^{\prime}}=2\sqrt{1-\eta_{\rm rad}/2}\xi^{\prime} a_0^3f(\xi^{\prime},a_0)~,
\label{eta-3}
\eeq
with an approximate solution 
\beq
\eta_{\rm rad}\approx 2\xi^{\prime} a_0^3f(\xi^{\prime},a_0)~,
\label{eta-4}
\eeq
which employs the fact that $\eta_{\rm rad}/2\ll 1$ up to very high values of $a_0$; in particular, at $a_0=750$ which was at the limit of our numerical calculation, $\eta_{\rm rad}/2\approx 0.11$.
In the limiting cases of weak and strong fields the integral in (\ref{fa}) can be solved analytically giving
\beq
\eta_{\rm rad}\approx\frac25\xi a_0^3~,~~~~a_0\ll a_{\rm cr}
\label{eta-small}
\eeq
and 
\beq
\eta_{\rm rad}\to 0.78~,~~~~a_0\to\infty~.
\label{eta-high}
\eeq
The latter number is obtained directly from (\ref{eta-3}), as the approximation $\eta_{\rm rad}/2\ll 1$ is no longer valid in this limit.
As is clearly seen from (\ref{eta-small}) in the limit of low intensities (which in practice means the laser field amplitude up to $a_0\simeq 400$), the field attenuation inside the plasma layer leads to further suppression of the radiation losses by a factor $\simeq 0.4$.

Note that the above results remain rather robust with respect to a particular model for the electron density and field distribution in the emitting layer.
As our simulations show, while the distribution of the electron density follows qualitatively that of Fig.3, the one for ions appears by far more complicated.
However, the only feature of the ion density distribution we practically use for the analytic modeling is that there is a significant number of ions to the left from the sharp electron density profile.
These ions, independently of the spatial shape of their distribution, create a quasistatic field $E_d$ which equilibrates the laser light pressure.
The electron density profile can also be chosen in different forms, and that given by Eq.(\ref{ne}) is not unique.
The only essential point is that both the electron density and the laser field amplitude drop down on the length $\ell_s$ which considerably reduces the effective value of $a_0$ particularly in the low-field regime, $a_0<a_{\rm cr}$.
The value of $\ell_s$ itself is also not of crucial significance as it enters the equations in the form of $N_x=n_{p0}\ell_s$.
Comparing the values of the area density $N_x$ calculated from (\ref{Nx}) and extracted from the simulation we found a reasonably good agreement: for $a_0=400$ the simulation and Eq.(3) give $N_x^{\mathrm{sim}}=(1.3\div 1.5)\cdot10^{19} {\rm cm}^{-2}$, and 
$N_x^{\mathrm{model}}=1.8\cdot 10^{19} {\rm cm}^{-2}$ correspondingly; for $a_0=500$ these numbers are $N_x^{\mathrm{sim}}=(1.4\div 1.7)\cdot10^{19} {\rm cm}^{-2}$, and $N_x^{\mathrm{model}}=2.2\cdot 10^{19} {\rm cm}^{-2}$.

A similar suppression effect emerges due to the laser amplitude dependence on the transverse coordinate and time.
Assuming that the dimensionless laser amplitude in the focal waist possess axial symmetry 
\beq
a(r,t)=a_0g(r/r_0,ct/r_L)
\label{art}
\eeq
and integrating the radiation power over the transverse coordinate and time we obtain that the function $f(\xi^{\prime},a_0)$ in (\ref{eta-4}) is replaced by the factor
\beq
S(\xi^{\prime},a_0)=\frac{\int g^5(\rho,\tau)f(\xi^{\prime},a_0g(\rho,\tau))d\rho d\tau}{\int g^2(\rho,\tau)d\rho d\tau}~.
\label{S1}
\eeq
where $\rho=(r/r_0)^2$ and $\tau=ct/r_L$.
Finally
\beq
\eta_{\rm rad}=2\xi^{\prime}a_0^3 S(\xi^{\prime},a_0)
\label{eta-5}
\eeq
apparently leading to additional suppression of the convergence efficiency.
For the supergaussian pulse (\ref{a-rxt}) used in the PIC simulations
\beq
S(\xi^{\prime},a_0)=\frac{1}{2^{1/4}\sqrt{\pi}\Gamma(1/4)}\int\limits_0^1\frac{dy_1}{y_1\sqrt{-\ln y_1}}
\int\limits_0^{y_1}\frac{dy_2y_2^4}{(-\ln (y_2/y_1))^{3/4}}f(\xi^{\prime},a_0y_2)~.
\label{S2}
\eeq
In the strong field limit $f(a)\sim 1/a^3$ so that the integrands in (\ref{S1}) are proportional one to another, leaving the limit (\ref{eta-high}) unchanged.
Instead, in the weak-field limit $f\approx 1/5$, which gives for (\ref{S2}) $S(a_0\ll a_{\rm cr})\approx 2^{3/4}/5^{7/4}$, and consequently $\eta_{\rm rad}\approx 0.20\xi^{\prime} a_0^3$. 
The resulting dependence $\eta_{\rm rad}(a_0)$ calculated for a supergaussian pulse (\ref{a-rxt}) along (\ref{eta-5}) and (\ref{S2}) is shown on Fig.\,1 by a solid red line and demonstrates an impressive improvement of (\ref{eta-new}): in the interval of intensities $a_0=400\div8000$ the calculated values do not deviate from the PIC result by more than 20\%.
Residual discrepancies may largely be ascribed to the fact that (\ref{V-hb}) tends to overestimate the actual recession velocity since complete reflection is assumed. Notice that radiation losses also contribute to decrease the reflectivity $R=1-\eta_{\rm rad}$ and hence reduce the recession velocity, which is principle may create a positive feedback for the enhancement of radiation emission. However, since $\eta_{\rm rad}$ is quite smaller than unity, these effects appear not to play a significant role.

\section{Extension of the classical regime of interaction towards higher intensities}

Although the radiation losses appear high compared to those in the ``light sail'' regime, their significant relative suppression caused by the RFF leads to a specific freezing of the electron lateral motion, so that the relativistic $\gamma$-factor grows much slower (\ref{gamma-new}) than in the perturbative domain $a_0\ll a_{\rm cr}$ where the RFF is negligible.
This in turn shifts the border between the classical and the quantum regime of interaction to considerably higher intensities.
The significance of QED effects is determined by the value of the relativistically invariant quantum parameter 
\beq
\chi=\frac{e\hbar}{m^3c^4}\sqrt{-(F^{\mu\nu}p_{\nu})^2}~,
\label{chi}
\eeq
where $F^{\mu\nu}$ is the electromagnetic field tensor and $p^{\nu}$ is the 4-momentum vector.
The value of (\ref{chi}) can be easily expressed via the parameters in the reference frame moving with $v_x=v_{\rm HB}$ where 
\beq
p^{\mu}=mc\gamma'(1,0,\sin(\varphi-\theta'),-\cos(\varphi-\theta'))~,
\label{p-mu}
\eeq
(see Eq.(\ref{v})).
Calculating the tensor $F^{\mu\nu}$ for the fields (\ref{EH}) and taking into account Eqs.(\ref{z})--(\ref{perp}), we obtain for (\ref{chi})
\beq
\chi=\frac{3}{2\alpha}\xi'(\gamma')^2~.
\label{chi-1}
\eeq
In the weak field regime $a_0\ll a_{\rm cr}$ this gives for 
$v_{\rm HB}\ll c$ a quantum parameter $\chi=(3/2\alpha)\xi a_0^2$, so that $\chi\simeq 0.1$ already at $a_0\approx 200$.
Recent work has shown that quantum quenching of radiation losses may be already significant at such modest values of $\chi$ \cite{zhangNJP15}.

However, in our particular conditions, due to (a) RF induced suppression in the growth of $\gamma'$ (\ref{gamma-new}) and (b) reduction of $\xi'$ with increasing of $v_{\rm HB}$ a further increase in the laser intensity results in a very slow growth of $\chi$ starting from $a_0\simeq a_{\rm cr}$.
In the strong field limit, $a_0\to\infty$ the hole boring velocity approaches the speed of light, and the parameter 
\beq
\xi'=\xi\sqrt{\frac{1-v_{\rm HB}/c}{1+v_{\rm HB}/c}}\simeq\frac{\xi}{\sqrt{2a_0}A^{1/4}}~,
\label{xi-new}
\eeq
where $A=Zn_cm_e/An_em_p$ (see Eq.(\ref{V-hb})). 
For parameters of our simulation $A\approx 3\cdot 10^{-6}$.
This results in the asymptotic value of the quantum parameter
\beq
\chi_{\infty}\approx\frac{3}{2\alpha}\bigg(\frac{a_0\xi^2}{2\sqrt{A}}\bigg)^{1/4}~.
\label{chi-infty}
\eeq
In the range $a_0=200 \div 800$ the value of $\chi_{\infty}$ increases from $0.092$ to $0.436$, and even for an ``extreme'' amplitude  of $a_0=2000$ we obtain $\chi\approx 0.636$, showing that the onset of a full radiation-dominated regime is prevented. 
In addition, with regards to the interaction geometry investigated in our case, these estimates neglect the screening of the laser field in the ``skin'' layer (Sec.\ref{sec:inhom}) from which most of the radiation is generated.   

This allows us to predict that, for the specific interaction geometry of CP pulses and thick overdense targets, QED effects will be strongly quenched compared to the case when the laser pulse and electron bunch counter-propagate or at least the longitudinal electron velocity $v_x\simeq 0$ in the laboratory frame.

\section{Conclusions}

In conclusion, we have presented a self-consistent analytic model for the interaction of superintense circularly polarized laser pulses with thick plasma in the hole boring regime.
The inclusion of the RFF along the lines of Zeldovich's work \cite{zeld} allowed calculating the conversion efficiency of the laser energy into high frequency radiation in the wide range of intensities.
After accounting the effects of (a) the global hole boring motion of the plasma and (b) of the laser field inhomogeneity in space and time, our result demonstrated a good quantitative agreement with the outcome of the PIC simulation.
Note that despite of its analytic simplicity the model is robust with respect to assumptions on the particular shape of electron and ion density distributions in the radiating layer.
The effect of the RFF, in combination with the factors (a) and (b), results in a much slower (compared to predictions made in \cite{ournjp}) increase of the conversion efficiency with the laser intensity, so that $\eta\approx 0.25$ at $I_L=3\cdot 10^{24}$W/cm$^2$.
Consequently, the quantum parameter also grows only slowly with increasing of the laser intensity, $\chi\sim a_0^{1/4}$, which may lead to quantum effects not to dominate even at the highest intensities we considered. This prediction may be tested by simulations with QED effects included.

\section*{Acknowledgements}
Authors acknowledge fruitful discussions with S.V. Bulanov, A.M. Fedotov, E.G. Gelfer, G. Korn, V.T. Tikhonchuck, and S. Weber.
SVP acknowledges support of the MEPhI Academic Excellence Project (Contract No. 02.a03.21.0005) and of the Russian Foundation for Basic Research through Grant No.16-02-00963a. The development of numerical algorithms was supported by Russian Science Foundation through Grant No.16-11-10028. Numerical simulations were performed using the computing resources granted by the John von Neumann-Institut f\"ur Computing (Research Center J\"ulich) under the project HRO04.

\section*{References}

\hyphenation{Post-Script Sprin-ger}
\providecommand{\newblock}{}


\begin{thebibliography}{10}
\expandafter\ifx\csname url\endcsname\relax
  \def\url#1{{\tt #1}}\fi
\expandafter\ifx\csname urlprefix\endcsname\relax\def\urlprefix{URL }\fi
\providecommand{\eprint}[2][]{\url{#2}}

\bibitem{bulanov-rmp06}
Mourou G~A, Tajima T and Bulanov S~V 2006 Optics in the relativistic regime
  {\em Rev. Mod. Phys.\/} {\bf 78}(2) 309--371

\bibitem{dipiazza-rmp12}
Di~Piazza A, M\"uller C, Hatsagortsyan K~Z and Keitel C~H 2012 Extremely
  high-intensity laser interactions with fundamental quantum systems {\em Rev.
  Mod. Phys.\/} {\bf 84}(3) 1177--1228

\bibitem{narozhny-cp15}
Narozhny N and Fedotov A 2015 Extreme light physics {\em Contemporary
  Physics\/} {\bf 56} 249--268

\bibitem{sokolov-jetp09}
Sokolov I~V 2009 Renormalization of the {Lorentz-Abraham-Dirac} equation for
  radiation reaction force in classical electrodynamics {\em Journal of
  Experimental and Theoretical Physics\/} {\bf 109} 207--212 ISSN 1090-6509

\bibitem{zotev-pop16}
Zot'ev D~B 2016 Critical remarks on {Sokolov}'s equation of the dynamics of a
  radiating electron {\em Physics of Plasmas\/} {\bf 23} 093302

\bibitem{landau-lifshitz-RR}
Landau L~D and Lifshitz E~M 1975 {\em The Classical Theory of Fields\/}
  (Elsevier, Oxford) chap~76 2nd ed

\bibitem{krivitskii-spu91}
Krivitskiĭ V~S and Tsytovich V~N 1991 Average radiation-reaction force in
  quantum electrodynamics {\em Soviet Physics Uspekhi\/} {\bf 34} 250

\bibitem{spohn-el00}
Spohn H 2000 The critical manifold of the {Lorentz-Dirac} equation {\em EPL
  (Europhysics Letters)\/} {\bf 50} 287

\bibitem{tamburini-njp10}
Tamburini M, Pegoraro F, {Di Piazza} A, Keitel C~H and Macchi A 2010 Radiation
  reaction effects on radiation pressure acceleration {\em New J. Phys.\/} {\bf
  12} 123005

\bibitem{vranic-cpc16}
Vranic M, Martins J, Fonseca R and Silva L 2016 Classical radiation reaction in
  particle-in-cell simulations {\em Computer Physics Communications\/} {\bf
  204} 141 -- 151 ISSN 0010-4655

\bibitem{cole-prx18}
Cole J~M, Behm K~T, Gerstmayr E, Blackburn T~G, Wood J~C, Baird C~D, Duff M~J,
  Harvey C, Ilderton A, Joglekar A~S, Krushelnick K, Kuschel S, Marklund M,
  McKenna P, Murphy C~D, Poder K, Ridgers C~P, Samarin G~M, Sarri G, Symes D~R,
  Thomas A~G~R, Warwick J, Zepf M, Najmudin Z and Mangles S~P~D 2018
  Experimental evidence of radiation reaction in the collision of a
  high-intensity laser pulse with a laser-wakefield accelerated electron beam
  {\em Phys. Rev. X\/} {\bf 8}(1) 011020

\bibitem{poder-prx17}
Poder K, Tamburini M, Sarri G, Di~Piazza A, Kuschel S, Baird C~D, Behm K,
  Bohlen S, Cole J~M, Corvan D~J, Duff M, Gerstmayr E, Keitel C~H, Krushelnick
  K, Mangles S~P~D, McKenna P, Murphy C~D, Najmudin Z, Ridgers C~P, Samarin
  G~M, Symes D~R, Thomas A~G~R, Warwick J and Zepf M 2018 Experimental
  signatures of the quantum nature of radiation reaction in the field of an
  ultraintense laser {\em Phys. Rev. X\/} {\bf 8}(3) 031004

\bibitem{macchi-p18}
Macchi A 2018 Viewpoint: Intense laser sheds light on radiation reaction {\em
  Physics\/} {\bf 11} 13

\bibitem{andersen-prd12}
Andersen K~K, Esberg J, Knudsen H, Thomsen H~D, Uggerhoj U~I, Sona P, Mangiarotti A, Ketel T~J, Dizdar A, Ballestrero S 2012Experimental investigations of synchrotron radiation at the onset of the quantum regime {\em Phys. Rev. D} {\bf 86}(7) 072001

\bibitem{wistisenNC18}
Wistisen T~N, Di Piazza A, Knudsen V and Uggerh{\o}j U I 2018 Experimental evidence of quantum radiation reaction in aligned crystals {\em Nature Comm.\/} {\bf 9} 795

\bibitem{naumova-prl09}
Naumova N, Schlegel T, Tikhonchuk V~T, Labaune C, Sokolov I~V and Mourou G 2009
  Hole boring in a {DT} pellet and fast-ion ignition with ultraintense laser
  pulses {\em Phys. Rev. Lett.\/} {\bf 102}(2) 025002

\bibitem{schlegel-pop09}
Schlegel T, Naumova N, Tikhonchuk V~T, Labaune C, Sokolov I~V and Mourou G 2009
  Relativistic laser piston model: Ponderomotive ion acceleration in dense
  plasmas using ultraintense laser pulses {\em Phys. Plasmas\/} {\bf 16} 083103

\bibitem{capdessus-pop14}
Capdessus R, Lobet M, {d'Humi\`eres} E and Tikhonchuk V~T 2014 $\gamma$-ray
  generation enhancement by the charge separation field in laser-target
  interaction in the radiation dominated regime {\em Phys. Plasmas\/} {\bf 21}
  123120

\bibitem{capdessus-pre15}
Capdessus R and McKenna P 2015 Influence of radiation reaction force on
  ultraintense laser-driven ion acceleration {\em Phys. Rev. E\/} {\bf 91}(5)
  053105

\bibitem{nerush-ppcf15}
Nerush E~N and Kostyukov I~Y 2015 Laser-driven hole boring and gamma-ray
  emission in high-density plasmas {\em Plasma Phys. Contr. Fusion\/} {\bf 57}
  035007

\bibitem{ournjp}
Liseykina T~V, Popruzhenko S~V and Macchi A 2016 Inverse {Faraday} effect
  driven by radiation friction {\em New J. Phys.\/} {\bf 18} 072001

\bibitem{delsorbo-njp18}
Del Sorbo D Blackman D~R, Capdessus R, Small K,Slade-Lowther C, Luo W, Duff M~J, Robinson A~P~L, McKenna P , Sheng Z-M, Pasley J and Ridgers C~P 2018 Efficient ion acceleration and dense electron–positron plasma creation in ultra-high intensity laser-solid interactions {\em New J. Phys.\/} {\bf 20} 033014

\bibitem{tamburiniPRE12}
Tamburini M, Lyseykina T V, Pegoraro F, and Macchi A , 2012 Radiation Pressure Dominant Acceleration: Polarization and Radiation Reaction Effects and Energy Increase in Three Dimensional Simulations {\em Phys. Rev. E} {\bf 85} 016407

\bibitem{zhangNJP15}
Zhang P, Ridgers C~P, and Thomas A~G~R 2015 The effect of nonlinear quantum electrodynamics on relativistic transparency and laser absorption in ultra-relativistic plasmas {\em New J. Phys.} {\bf 17} 043051

\bibitem{zeld}
Zeldovitch Y~B 1975 Interaction of free electrons with electromagnetic radiation {\em Soviet Physics Uspekhi\/} {\bf 18} 79


\bibitem{robinson}
Robinson A~P~L, Gibbon P, Zepf M, Kar S, Evans R~G and Bellei C 2009
  Relativistically correct hole-boring and ion acceleration by circularly
  polarized laser pulses {\em Plasma Phys. Contr. Fusion\/} {\bf 51} 024004

\bibitem{esirkepovPRL04}
Esirkepov T, Borghesi M, Bulanov S~V, Mourou G and Tajima T 2004 Highly
  efficient relativistic-ion generation in the laser-piston regime {\em Phys.
  Rev. Lett.\/} {\bf 92} 175003

\bibitem{macchiNJP10}
Macchi A, Veghini S, Liseykina T~V and Pegoraro F 2010 Radiation pressure
  acceleration of ultrathin foils {\em New J. Phys.\/} {\bf 12} 045013

\bibitem{landau-lifshitz-Prad}
Landau L~D and Lifshitz E~M 1975 {\em The Classical Theory of Fields\/}
  (Elsevier, Oxford) chap~78 2nd ed

\bibitem{macchi-prl05}
Macchi A, Cattani F, Liseykina T~V and Cornolti F 2005 Laser acceleration of
  ion bunches at the front surface of overdense plasmas {\em Phys. Rev.
  Lett.\/} {\bf 94}(16) 165003

\bibitem{kost-pop16}
Kostyukov I~Y and Nerush E~N 2016 Production and dynamics of positrons in
  ultrahigh intensity laser-foil interactions {\em Physics of Plasmas\/} {\bf
  23} 093119

\bibitem{bulanovjrPRL10}
Bulanov S~S, Esirkepov T~Z, Thomas A~G~R, Koga J~K and Bulanov S~V 2010
  Schwinger limit attainability with extreme power lasers {\em Phys. Rev.
  Lett.\/} {\bf 105}(22) 220407

\end{thebibliography}

\end{document}